\documentclass[useAMS,usenatbib,a4paper]{mn2e}

\usepackage{graphicx}
\usepackage{times}
\usepackage{amsmath}
\usepackage{natbib}
\usepackage[usenames]{color}

\newcommand{\sref}[1]{Section~\ref{#1}}
\newcommand{\fref}[1]{Fig.~\ref{#1}}

\newcommand{\tref}[1]{Table~\ref{#1}}

\begin{document}

\title[The puzzle of HSC~J115252+004733]{A new quadruple gravitational lens from the Hyper Suprime-Cam Survey: the puzzle of HSC~J115252+004733}

\author[More et al.]{
Anupreeta More,$^{1}$\thanks{anupreeta.more@ipmu.jp} 
Chien-Hsiu Lee,$^{2}$ 
Masamune Oguri,$^{1,3,4}$ 
Yoshiaki Ono,$^{5}$
\newauthor{ 
Sherry H. Suyu,$^{6,7}$ 
James H. H. Chan,$^{7,8}$ 
John D. Silverman,$^{1}$ 
Surhud More,$^{1}$ 
}
\newauthor{
Andreas Schulze,$^{1}$ 
Yutaka Komiyama,$^{9,10}$
Yoshiki Matsuoka,$^{9}$
Satoshi Miyazaki,$^{9,10}$
}
\newauthor{
Tohru Nagao,$^{11}$
Masami Ouchi,$^{5}$
Philip J. Tait,$^{12}$
Manobu M. Tanaka,$^{13}$
}
\newauthor{
Masayuki Tanaka,$^{9}$
Tomonori Usuda$^{9,10}$
and
Naoki Yasuda$^{1}$
}
\medskip
\\
$^{1}$ Kavli Institute for the Physics and Mathematics of the Universe (Kavli IPMU, WPI), University of Tokyo, Chiba 277-8583, Japan \\
$^{2}$ Subaru Telescope, National Astronomical Observatory of Japan, 650 North Aohoku Place, Hilo, HI 96720, USA\\
$^{3}$ Research Center for the Early Universe, University of Tokyo, 7-3-1 Hongo, Bunkyo-ku, Tokyo 113-0033, Japan\\
$^{4}$ Department of Physics, University of Tokyo, 7-3-1 Hongo, Bunkyo-ku, Tokyo 113-0033, Japan\\
$^{5}$ Institute for Cosmic Ray Research, The University of Tokyo, 5-1-5 Kashiwa-no-Ha, Kashiwa City, Chiba, 277-8582, Japan\\
$^{6}$ Max-Planck-Institut f{\"u}r Astrophysik, Karl-Schwarzschild-Str.~1, 85748 Garching, Germany\\
$^{7}$ Institute of Astronomy and Astrophysics, Academia Sinica, P.O. Box 23-141, Taipei 10617, Taiwan\\
$^{8}$ Department of Physics, National Taiwan University, Taipei 10617, Taiwan\\
$^{9}$ National Astronomical Observatory of Japan, 2-21-1 Osawa, Mitaka, Tokyo 181-8588, Japan\\
$^{10}$ SOKENDAI (The Graduate University for Advanced Studies), Osawa, Mitaka, Tokyo 181-8588, Japan\\
$^{11}$ Research Center for Space and Cosmic Evolution, Ehime University, Bunkyo-cho 2-5, Matsuyama, Ehime 790-8577, Japan\\
$^{12}$ Subaru Telescope, 650 North A’ohoku Place, Hilo, HI 96720, USA\\
$^{13}$ High Energy Accelerator Research Organization, Institute of Particle and Nuclear Studies, KEK, 1-1 Oho, Tsukuba, Ibaraki 305-0801, Japan\\
}

\maketitle

\begin{abstract}

We report the serendipitous discovery of a quadruply lensed source
at $z_{\rm s}=3.76$, HSC~J115252+004733, from the Hyper Suprime-Cam
(HSC) Survey. The source is lensed by an early-type galaxy at $z_{\rm
l}=0.466$ and a satellite galaxy. Here, we investigate the properties of the
source by studying its size and luminosity from the imaging and the
luminosity and velocity width of the Ly-$\alpha$ line from the spectrum.
Our analyses suggest that the source is most probably a low-luminosity
active galactic nucleus (LLAGN) but the possibility of it being a
compact bright galaxy (e.g., a Lyman-$\alpha$ emitter or Lyman Break
Galaxy) cannot be excluded. The brighter pair of lensed images appears
point-like except in the HSC $i$-band (with a seeing $\sim0\farcs5$).
The extended emission in the $i$-band image could be due to the host
galaxy underneath the AGN, or alternatively, due to a highly compact
lensed galaxy (without AGN) which appears point-like in all bands except
in $i$-band. We also find that the flux ratio of the brighter pair of
images is different in the Ks-band compared to optical
wavelengths. Phenomena such as differential extinction and intrinsic
variability cannot explain this chromatic variation. While microlensing
from stars in the foreground galaxy is less likely to be the cause, it
cannot be ruled out completely. If the galaxy hosts an AGN, then this
represents the highest redshift quadruply imaged AGN known to date,
enabling study of a distant LLAGN. Discovery of this unusually compact
and faint source demonstrates the potential of the HSC survey.

\end{abstract}


\section{Introduction}
\label{sec:intro}

Strong gravitational lensing is a powerful astrophysical and
cosmological tool. Gravitational lenses act as natural telescopes to
provide a sneak peek at some of the most distant and faintest galaxies
and quasars in the Universe.

Distant galaxies are mostly comprised of star-forming galaxies
(SFGs) often identified in observations as Lyman Break Galaxies (LBGs)
or Lyman-$\alpha$ emitters (LAEs). The magnification due to gravitational
lensing is critical to study the faint end of the distant LAE and LBG
populations at high angular resolution. Lensing has enabled several
studies such as sources of cosmic reionization
\citep[e.g.,][]{Stark2007,Atek2015}, understanding the physical properties
of LAEs/LBGs at high redshifts $z=6-7$
\citep[e.g.,][]{Egami2005,Bradley2008,Huang2016}, their molecular gas and
inter-stellar medium (ISM) properties
\citep[e.g.,][]{Riechers2010,Rawle2014} as well as their abundances
\citep[e.g.,][]{Bradley2014,Schmidt2016}.

Lensed quasars have also proven to be useful as both cosmological and
astrophysical probes. For example, the time delays provide a direct way
to measure cosmological parameters \citep[e.g.,][]{Suyu2016} and source
reconstruction of lensed quasars and their host gives a direct view on
the co-evolution of quasars and host galaxies \citep[e.g.,][]{Peng2002,
Rusu2016}.  Systematic searches for lensed quasars have successfully
found over a hundred lens systems, in the radio
\citep[e.g.,][]{Myers2003, Browne2003}, in the optical
\citep[e.g.,][]{Oguri2006,Inada2012,More2016} as well as other
multi-wavelength regimes \citep[e.g.,][]{Jackson2012}.  On galaxy
scales, lensed quasars are typically doubly imaged (``doubles'') or
quadruply imaged (``quads''). Most of the lensed quasar systems
discovered to date are doubles. For example, a sample of thirteen lensed
quasars recently discovered by \citet{More2016} from the Sloan Digital
Sky Survey (SDSS) -III  are all doubles.  Nonetheless, quads with their
two additional images provide additional astrophysical information on
the foreground lens and the background source.  Finding more quad quasar
lenses is thus of tremendous value to the community given the small
number\footnote{According to the Master Lens Database
(\texttt{http://admin.masterlens.org/index.php}), from a collection of
over 115 lensed quasars, about 30 systems are known to be quads.} of
currently known quads.

In this paper, we report the discovery of the quad lens
HSC~J115252+004733 (henceforth referred to as HSC~J1152+0047)
from the Hyper Suprime-Cam (HSC) Survey. The HSC Survey is a Subaru
Strategic Program (SSP) using the newly installed HSC
\citep{Miyazaki2012} instrument on the Subaru 8.2-m telescope. The
survey consists of three layers (wide, deep, ultradeep), where the wide
layer is expected to cover $\sim1400$ deg$^2$ in $grizY$-bands down to a depth of
$r\sim26$. The HSC data are processed with hscPipe, which is
derived from the LSST software pipeline \citep{Ivezic2008, Axelrod2010,
Juric2015}, and are calibrated using the Pan-STARRS1 data
\citep{Tonry2012, Schlafly2012, Magnier2013}.

Our paper is organized as follows. In \sref{sec:phot}, we describe the
discovery of HSC~J1152+0047 and the multi-wavelength imaging data on
this lens system. We describe the spectroscopic follow-up in
\sref{sec:spec}. The lens mass modeling is described in
\sref{sec:massmodel}.  In \sref{sec:res}, we derive the properties of
the delensed source. In \sref{sec:disc}, we compare our source with
other distant galaxies and quasars to understand its nature and discuss
the cause of chromatic variation in flux ratios.  We present our
conclusions in \sref{sec:conc}. Magnitudes quoted in this paper are in
AB magnitudes unless otherwise stated. The terms active galactic nuclei
(AGN) and quasar may be used interchangeably in the text.  We used the
following cosmological parameters wherever necessary $\Omega_m=0.308$,
$h=0.678$, $\Omega_k=0$ \citep{PlanckCollaborationandAde2016}.

\section{Imaging Data of HSC~J1152+0047}
\label{sec:phot}

\begin{figure}
\begin{center}
\includegraphics[scale=0.70]{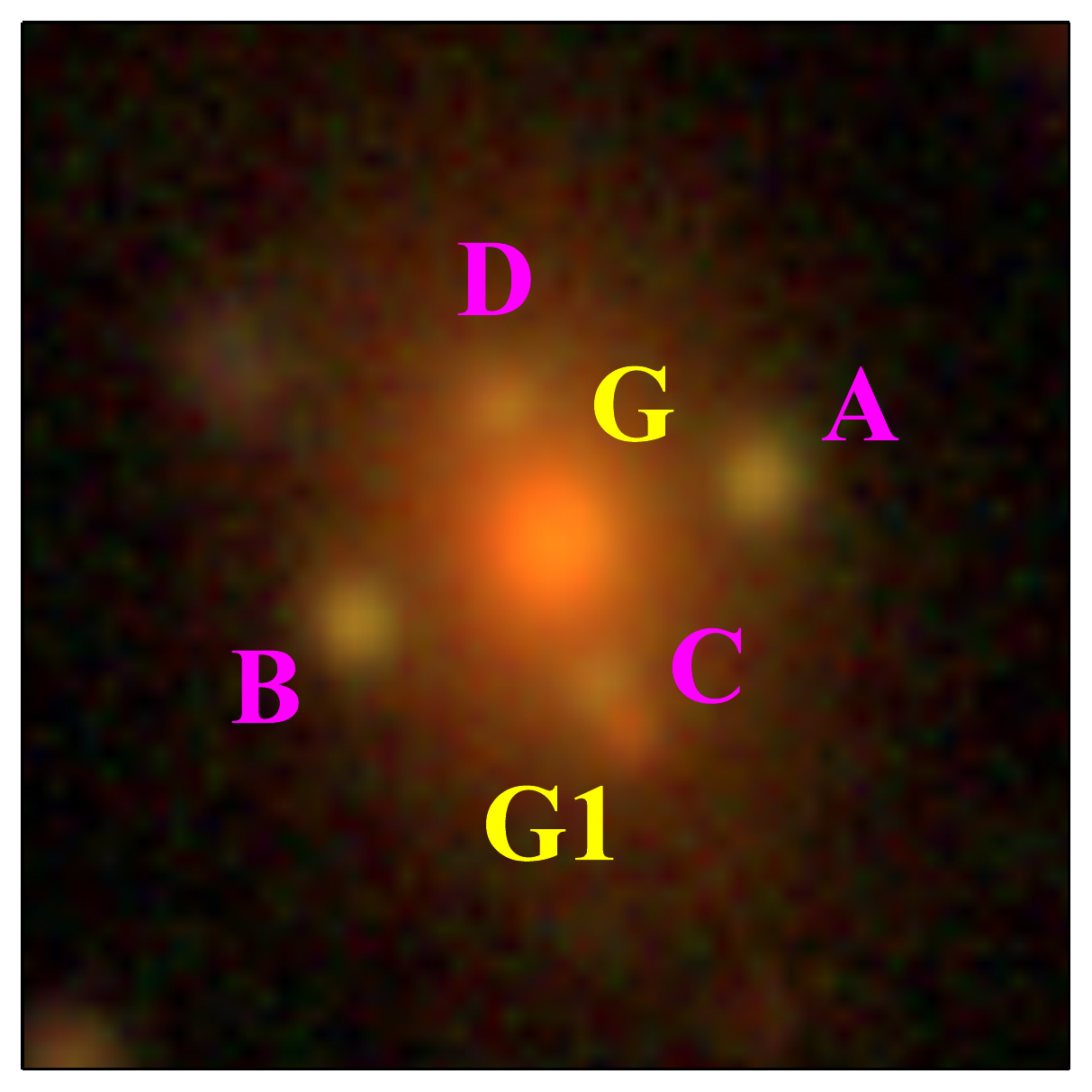}
\caption{\label{fig:colimg} Color ($gri$) composite of HSC~J1152+0047
showing the four blue lensed images (A, B, C and D) in an Einstein-cross
configuration. Apart from the central lensing galaxy (G), a companion
galaxy (G1, close to image C) is probably contributing to lensing. The
image is 10\arcsec~on the side. North is up and East is
left.}
\end{center}
\end{figure}

\begin{figure*}
\includegraphics[scale=0.55]{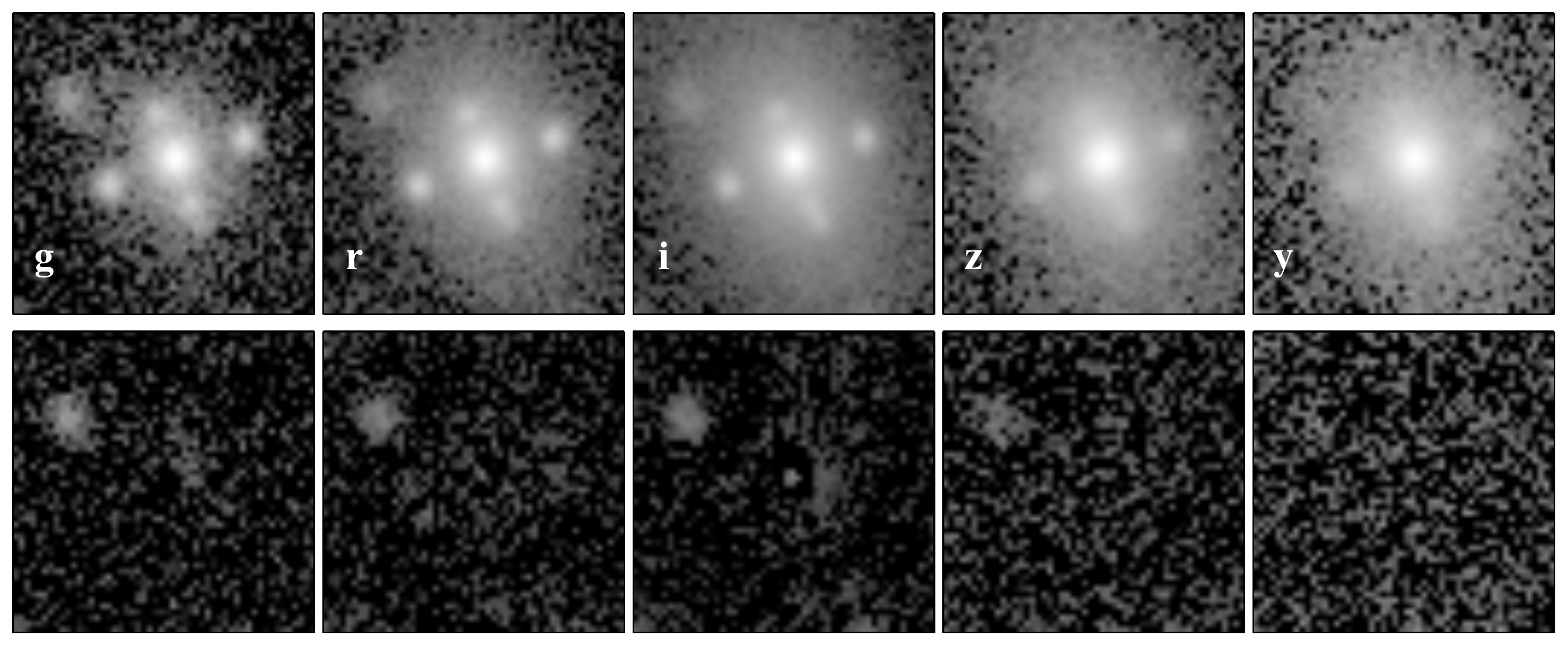}
\caption{\label{fig:galfit} {\sc Galfit} modeling results in HSC~$g r
  i z Y$-bands. Top panels show the HSC images whereas the bottom
  panels show {\sc Galfit} model subtracted residual images in the
  respective bands. Images are 9\arcsec on the side.}
\end{figure*}

The lens system, HSC~J1152+0047, was recently discovered serendipitously
during the visual inspection of data from the HSC Wide (internal data
release $\sim80$~sq.~deg., S15A).  HSC~J1152+0047 consists of the main lens galaxy (G)
with four lensed images (A, B, C and D) in a cross configuration (see
\fref{fig:colimg}).  A second galaxy (G1) located very close to image
C, is probably a satellite of the main lens galaxy G and is likely to
perturb the lens potential. 

We measured the relative positions and photometry of the lens galaxies
and the lensed images from the HSC imaging using {\sc galfit}
\citep{Peng2002}.  In \fref{fig:galfit}, we show all of the HSC bands
and  the model-subtracted residual images for each band, respectively.
The seeing in the HSC-$g,r,i,z$ and $Y$ images is found to be $0\farcs55$,
$0\farcs46$, $0\farcs45$, $0\farcs60$, and $0\farcs61$, respectively, as
per hscPipe. All of the four lensed images are fit with a point spread
function (PSF) model in all bands except in the $i$-band, where a Sersic
profile is better fit to images A and B. The lens galaxy G is fit with a double
Sersic model and the companion galaxy G1 is fit with a PSF model. For
lens galaxy G, we used the $z$-band best-fit model as prior when fitting
the double Sersic model in other bands. The relative positions from
$z$-band and photometry in all bands along with errors from {\sc galfit}
are given in \tref{tab:phot}. We note that the colors ($g-r$ and $r-i$)
of the lensed images (see \tref{tab:phot}) are consistent with the
colors of a quasar at $z\sim4$ \citep{Richards2001}.

\begin{figure*}
\includegraphics[scale=0.43]{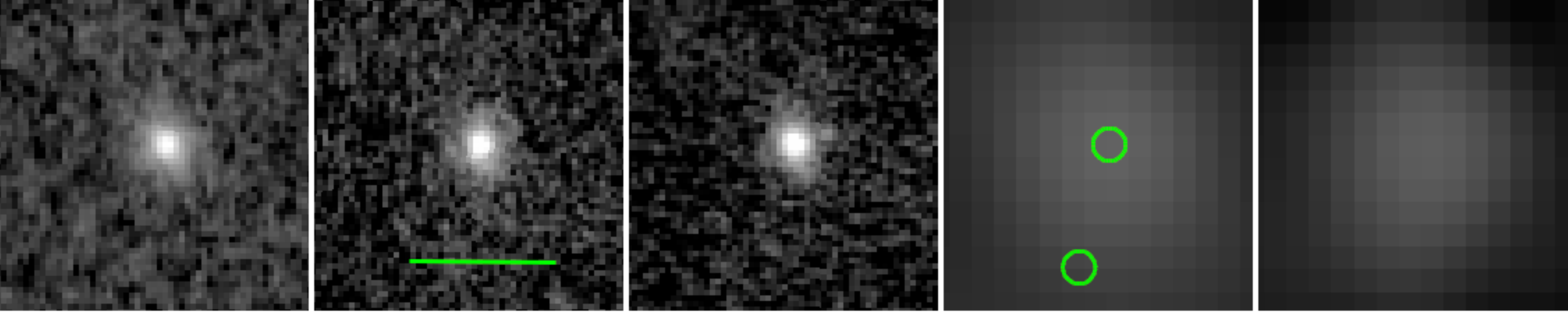}
\caption{\label{fig:ir} From left to right, NIR VIKING imaging in
Filters (limiting AB magnitudes) $J (22.1)$, $H (21.5)$, and
$K$s$(21.2)$ followed by WISE imaging in bands W1 and W2. The brighter
pair of lensed images are close to the detection limits of VIKING $J$
and $H$ imaging whereas they are detected in the $Ks$-band. W1 and W2
imaging show emission from the lens galaxy alone (see \sref{sec:phot}).
The two green circles mark the location of sources detected from
the WISE catalogs. North is up and East is left. The bar shows a scale
of 9\arcsec. }
\end{figure*}

We find that our system is detected in the near-infrared (NIR)
imaging taken by the VISTA Kilo-degree Infrared Galaxy survey
\citep[VIKING;][]{Edge2013}. The $JHK$s imaging obtained from the VISTA
Science Archive
(VSA)\footnote{\texttt{http://horus.roe.ac.uk/vsa/dboverview.html}} is
shown in \fref{fig:ir}. The brighter pair of lensed images shows hints
of presence in the $J$ and $H$ imaging, whereas they are well detected
in the $K$s-band. We fit a PSF model to the brighter pair of images and
a Sersic model to the galaxy G with {\sc galfit}. We use the 2MASS
$K$-band magnitude of a nearby star for the flux calibration of our
VIKING $K$s-band photometry.  The images A, B and galaxy G have
$K$s-band magnitudes of $19.31\pm0.10$, $20.28\pm0.27$ and
$15.23\pm0.14$ Vega mag, respectively where the reported errors are
from {\sc galfit}.

We also checked if the lensed source is detected in the Wide-field
Infrared Survey Explorer (WISE) imaging \citep{Wright2010}. While W1-
and W2-bands show emission at the center of HSC~J1152+0047, we cannot
confirm the presence of the lensed source due to source confusion owing
to the low resolution (6\arcsec). The source catalog from WISE reports
two detections. One of the sources coincides with the lens galaxy G
whereas the second source does not coincide with any of the lensed
images or galaxy G1 (see \fref{fig:ir}). The lens galaxy G has
$W1=15.18\pm0.04$ and $W2=15.04\pm0.08$ Vega mag, respectively. In
bands W3 and W4, there is no detection of any emission either from the
lens or the lensed source thus, providing a strong upper limit on source
flux. The sensitivities with a signal-to-noise ratio (SNR) of 5 for
W3 and W4 bands are 11.5 and 7.87~Vega mag, respectively.

\begin{figure}
\includegraphics[scale=0.6]{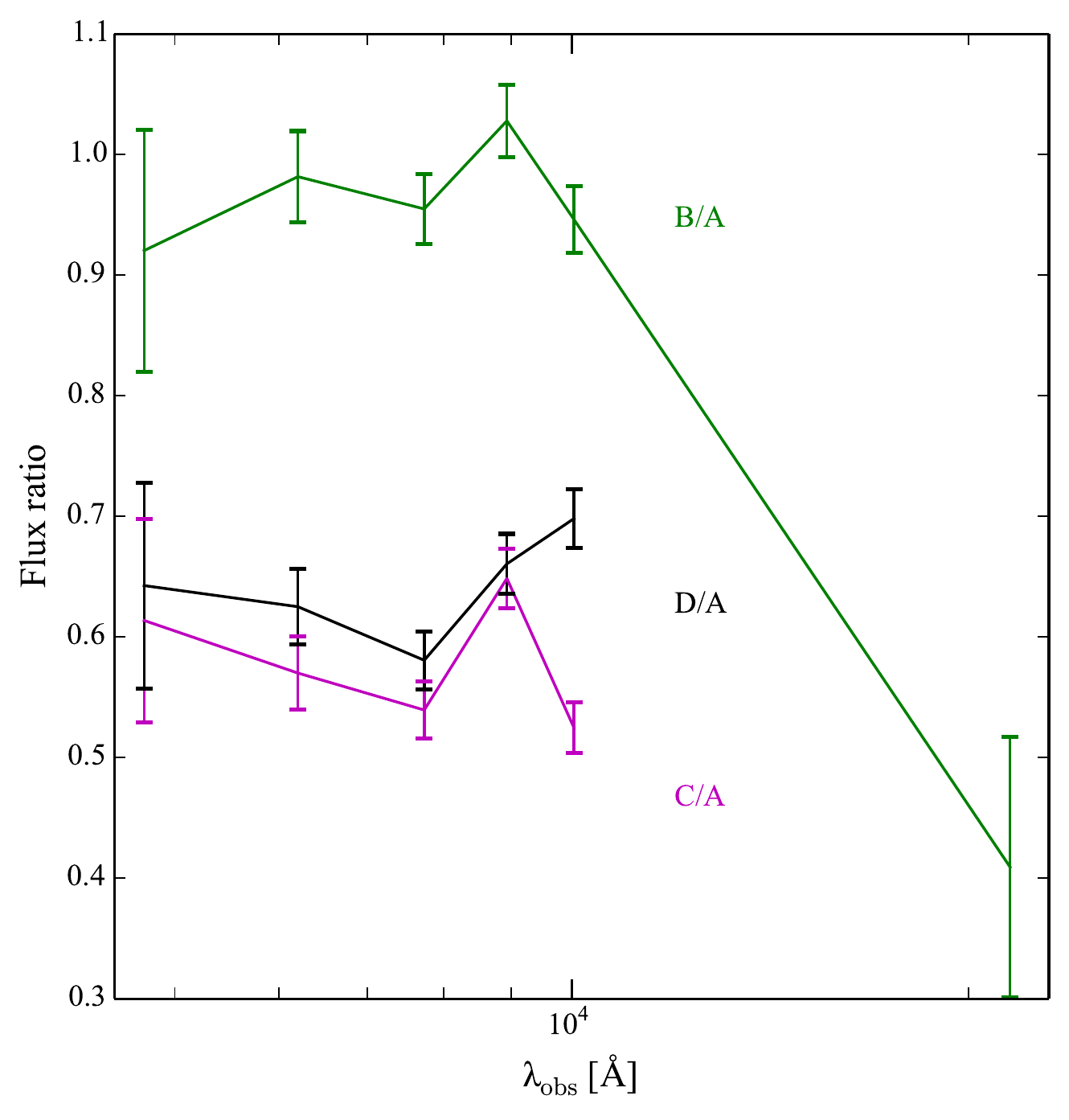}
\caption{\label{fig:fxrat} Flux ratios of the lensed images as  a
  function of wavelengths in the optical (HSC $grizY$-band, see also
  \tref{tab:phot}). For the brighter pair of images, we also include
the $K$s-band data in the NIR.}
\end{figure}

We show the flux ratios of images B, C, D with respect to image A in
\fref{fig:fxrat} for all of the HSC $grizY$-bands and VIKING $K$s-band.
All of the three flux ratios appear nearly uniform across the optical
bands with no strong evidence for differential reddening.  However, the
flux ratio of the brighter image pair (B/A$\sim 0.41\pm0.11$) in
$K$s-band (observed in year 2011) is discrepant with the flux ratio of
$0.95\pm0.02$ seen in the HSC data (observed in year 2015). In
\sref{sec:chrom}, we discuss various phenomena that are known to
introduce chromatic (i.e. wavelength dependent) variation in the flux
ratios of lensed images in the attempt to find plausible explanation for
this discrepancy.

Images C and D are detected in the optical bands only. The flux of image
C is likely contaminated by the emission from satellite galaxy (G1) in
the reddest ($Y$) band where G1 becomes more prominent (see
\fref{fig:fxrat} and \tref{tab:phot}). In the bluer bands, the fluxes of
images C and D are comparable as expected for this image configuration
from lensing.  Although the presence of G1 could affect the
magnification of image C, we do not detect any significant difference
between relative fluxes of C and D. 

\begin{table}
\begin{center}
\caption{\label{tab:phot}Relative astrometry and photometry of
HSC~J1152+0047 using {\sc galfit}.}
\begin{tabular}{crrrrrrr}
\hline\hline
Name & $\Delta x$ & $\Delta y$ & $g$ & $r$ & $i$ & $z$ & $Y$ \\
     &  err & err & err & err & err & err & err \\
\hline
A  & 	    0.0  &    0.0    &  24.23  &  23.10  &  22.82  &  22.82  &  22.80  \\  
   & 	     --  &     --    &   0.06  &   0.02  &   0.02  &   0.02  &   0.04  \\  
B  & 	$-$3.85  &  $-$1.34  &  24.32  &  23.12  &  22.87  &  22.79  &  22.86  \\  
   & 	   0.01  &     0.01  &   0.06  &   0.02  &   0.02  &   0.02  &   0.04  \\  
C  & 	$-$1.52  &  $-$1.87  &  24.76  &  23.71  &  23.49  &  23.29  &  23.50  \\  
   & 	   0.02  &     0.02  &   0.10  &   0.05  &   0.03  &   0.03  &   0.15  \\  
D  & 	$-$2.45  &     0.70  &  24.71  &  23.61  &  23.41  &  23.27  &  23.19  \\  
   & 	   0.01  &     0.01  &   0.10  &   0.05  &   0.03  &   0.03  &   0.07  \\  
G  & 	$-$1.96  &  $-$0.57  &  21.85  &  20.31  &  19.52  &  19.15  &  19.05  \\  
   & 	   0.01  &     0.01  &   0.12  &   0.05  &   0.05  &   0.02  &   0.03  \\  
G  & 	$-$2.04  &  $-$0.47  &  21.63  &  20.27  &  19.43  &  19.18  &  19.06  \\  
   & 	   0.01  &     0.02  &   0.22  &   0.03  &   0.12  &   0.03  &   0.03  \\  
G1 & 	$-$1.25  &  $-$2.36  &  25.87  &  24.43  &  23.62  &  23.22  &  23.01  \\  
   & 	   0.01  &     0.02  &   0.24  &   0.07  &   0.03  &   0.04  &   0.11  \\  
\hline
\end{tabular}
\flushleft{The position of image A is RA,Dec=(178.21722, 0.79271) deg.
The relative astrometry is from the $z$-band in units of arcsec. 
The positive directions of $\Delta x$ and $\Delta y$ are North and
West, respectively. For every object, the second row shows the errors
from {\sc galfit}. Except the $i$-band for images A and B where a
Sersic model is chosen, all of the lensed images are fit with a PSF
model. The magnitudes are corrected for Galactic extinction.}
\end{center}
\end{table}

\section{Spectroscopic observation of HSC~J1152+0047}
\label{sec:spec}

We obtained spectra of HSC~J1152+0047 on 2016 March 11 (Program
ID: GN-2016A-FT-7) with Gemini North Telescope through the Fast
Turnaround program \citep{Mason2014} with seeing around
$0\farcs6$-$0\farcs7$.  We used GMOS in the long-slit mode with
R400-G5305 grating and GG455 blocking filter.  The width of the slit was
set to $1\farcs0$. This provides a spectral resolution R=1918 and
wavelength coverage from 4000 to 8000~\AA.  Two perpendicular slit
configurations were used to acquire spectra of all 4 lensed images, as
well as the lens galaxy G. The brighter pair was observed for 15~min and
the fainter pair was observed for 45~min. The companion galaxy G1 is
aligned with the fainter pair and fell on the second slit but was too
faint to be detected. 

We carried out the data reduction using IRAF with the dedicated GMOS
package (v1.13). The spectra were bias subtracted, flat fielded and sky
subtracted. We used CuAr lamp to calibrate the wavelength. Because the
lensed image spectra are very faint and close to the spectra of the lens
galaxy ({$<$2\arcsec} for the brighter pair, and {$<$1\arcsec} for the
fainter pair), we manually select the aperture size and spectra position
to extract the lensed image spectra. This is done using the `gsextract'
routine in an interactive mode. For the fainter pair, the spectra are
dominated by the lens galaxy light. To avoid contamination from the
lens for the fainter pair of images, we scaled the red part
($>7000$~\AA) of the lens spectrum to match the spectrum of the fainter
images assuming that most the of redder part is dominated by the
emission from the lens. We then subtracted this normalized lens galaxy
spectrum from the spectra of the fainter images. We calibrate the flux
based on spectroscopic standard star. We estimate flux errors using the
variance at each pixel, which is calculated from the raw image, and
taking into account the effects of each reduction step using error
propagation, till extracting the one-dimensional spectra.  

\begin{figure*}
\begin{center}
\includegraphics[scale=0.57]{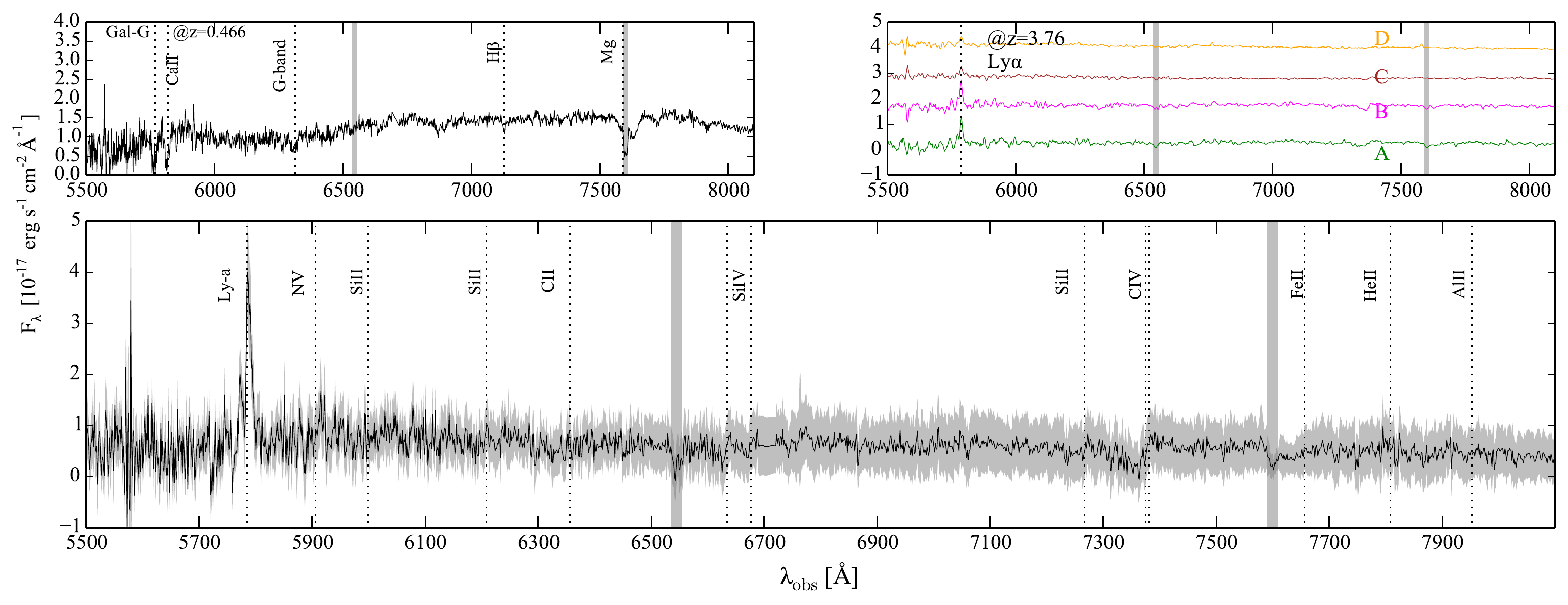}
\end{center}
\caption{\label{fig:spec} {\it Upper left:} The spectrum of the lensing
galaxy from our Gemini GMOS spectroscopic follow-up observation. The
lens  redshift is $z_{\rm l}=0.466$. {\it Upper right:} Individual
binned spectra of the lensed images A--D from the Gemini GMOS spectroscopy. All
the images have the Ly-$\alpha$ emission line redshifted to $z_{\rm
s}=3.76$.  {\it Lower:} The stacked spectrum of the lensed
source with commonly found emission and absorption lines labelled.
The error on the spectrum is shown with the shaded region (grey). The two
vertical bars (grey) shown in all panels indicate absorption features
probably due to telluric contamination.
} \end{figure*}

The spectra of the lens galaxy and the four lensed images are shown in
\fref{fig:spec}. The spectra of each of the lensed images in the figure
are smoothed with a Gaussian filter of 2.7~\AA~and vertical offsets are
added for illustrative purpose. The redshift of the lens galaxy is found
to be $z_{\rm l}=0.466\pm0.001$ through the identification of CaII~H and
K absorption lines near the 4000~\AA~break as expected for a typical
early-type galaxy. As mentioned earlier, we do not have a spectrum of G1
but our photometric redshift estimate ($z_{l{\rm G1}}=0.46\pm0.15$)
suggests that G1 is a gravitationally bound satellite of galaxy G.  The
similarity in the spectra of the four images confirms the strong lensing
nature of this system. The redshift of the lensed source is found to be
$z_{\rm s}=3.76\pm0.01$ based on the identification of the Ly-$\alpha$
emission line.  Also, the stacked spectrum shows other high 
ionization state lines e.g., NV in emission and C~IV in absorption which are rather
weak but their presence is consistent with the interpretation of the
stronger emission line as Ly-$\alpha$ (see \fref{fig:spec}).

\section{Lens Mass Modeling}
\label{sec:massmodel}

We used the lens modeling software {\sc glafic} \citep{Oguri2010} for
lens mass modeling. We started with $i$-band HSC image where
information from each pixel offer data constraints for the mass models.

The lens galaxies are each modeled with an isothermal density profile.
Galaxy G is allowed to have ellipticity whereas galaxy G1 is assumed
to have spherically symmetric mass distribution since G1 is less massive
and the contribution from the quadrupole moment of its mass distribution
is not expected to be significant.  While we allowed for external shear
to be present earlier in our models, we found that this was highly
degenerate with the ellipticity of G and did not improve the fit
significantly. Hence, we removed it for simplicity. Thus, our final lens
mass model consists of an isothermal ellipsoid (SIE) for G and an
isothermal sphere (SIS) for G1.

We tested different assumptions when modeling the light profile of the
lensed source. Our list of models comprised i) single PSF ii) single
Sersic and iii) PSF+Sersic. We found that a single PSF model left
significant residuals in the $i$-band image, as expected and a
PSF+Sersic model was not an improvement over the single Sersic
model. Therefore, our final best model for the lensed source is a
single Sersic component.

\begin{figure*}
\includegraphics{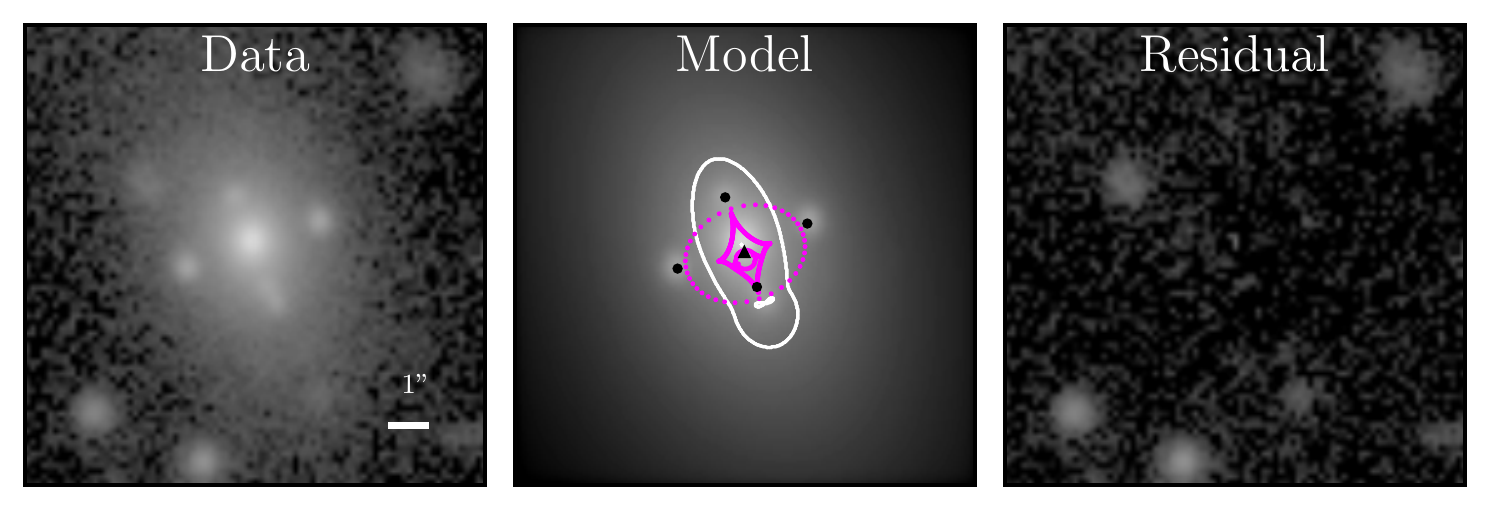}
\caption{\label{fig:massmod} HSC~$i$-band image ({\it left}), lens mass model
 where the background source is assumed to have a Sersic profile
({\it middle}) and the residual image ({\it right}). The middle panel
 also shows the positions of the lensed images (circles) and the
 source (triangle). The magenta contours show the caustics in the
 source plane and white contours mark the corresponding critical
 curves in the image plane indicating regions of extremely high
 magnification.} 
\end{figure*}

We run Monte Carlo Markov Chain (MCMC) using {\sc emcee}
\citep{ForemanMackey2013}. The model fitting includes pixels within a
$7\arcsec$ box centered on the lens (see left panel of
\fref{fig:massmod}). The medians of the posterior distributions of our
model parameters are given in \tref{tab:lmod}. We only report the source
magnitude and effective radius from our model in \tref{tab:lmod} since
other source parameters such as the ellipticity and position angle (PA)
are not well-constrained. Also, since the Sersic index $n$ was not
well-constrained we fixed it to $n=4$ that corresponds to the de
Vaucoleurs' profile. 

In \fref{fig:massmod}, we show the model image (middle panel) and the
residual image (right panel) after subtracting the model from the data.
We also show critical curves (white contours) and caustics (magenta
contours) marking the regions of infinite magnification in the middle
panel. We highlight the peak positions of the model-predicted lensed
images (circles) and the true source position (triangle).

Finally, we also constructed mass models by fitting to just the peak
positions and relative fluxes of the lensed images which were measured
with {\sc galfit}. We tested similar lens models as before except that
the source is assumed to be a point source since the data constraints
are limited. From this analysis, we determined a magnification factor
of $\mu=2.5-3$ for images A and B. We choose $\mu=2.5$\footnote{We
  note that choosing $\mu=3$ does not qualitatively change our
  conclusions about the source properties.} for the analysis presented
in \sref{sec:res}. 

\begin{table}
\begin{center}
\caption{\label{tab:lmod} Mass modeling results from fitting a lensed extended
source to $i$-band image.}
\begin{tabular}{ll}
\hline\hline
Parameters & Values (units) \\
\hline
Velocity dispersion (G)  & $280 \pm 10$ km~s$^{-1}$    \\ 
Ellipticity (G)          & $0.54 \pm 0.02$              \\ 
PA of Ellipticity (G)    & $19.1 \pm 0.5$  deg          \\ 
Velocity dispersion (G1) & $100 \pm 30$ km~s$^{-1}$     \\ 
True Source magnitude         & $23.9 \pm 0.1$             \\ 
True Source Effective radius  & $0.028 \pm 0.005$~arcsec     \\ 

\hline
\end{tabular}
\end{center}
\end{table}

\begin{table}
\begin{center}
\caption{\label{tab:meas} Properties of Ly-$\alpha$ for the brighter
pair of lensed images.}
\begin{tabular}{llll}
\hline\hline
Name & EW  & FWHM & $L \times 10^{42}$ \\
  & \AA & km~s$^{-1}$ & ergs~s$^{-1}$ \\
\hline
A & $16\pm1$ ($30$) & $540\pm40$ ($920$) & 3.2 (6.3) \\
B & $15\pm1$ ($26$) & $640\pm50$ ($890$) & 3.2 (6.3) \\
\hline
\end{tabular}
\flushleft{The values are given for the narrow component of
Ly-$\alpha$ and the values within parentheses are given for the broader
component. The EW is given in the rest-frame. The luminosities are delensed by
taking into account the magnification factor.}
\end{center}
\end{table}

\section{Results}
\label{sec:res}

In this section, we determine the properties of the lensed source based on
the modeling of the imaging and spectroscopic data. First, we analyze
the single most prominent emission line found in the spectra of the
lensed images. Next, we determine the true source magnitude and size to
decide whether the source is dominated by an AGN or not. 

\subsection{Ly-$\alpha$ emission}

\begin{figure}
\begin{center}
\includegraphics[scale=0.70]{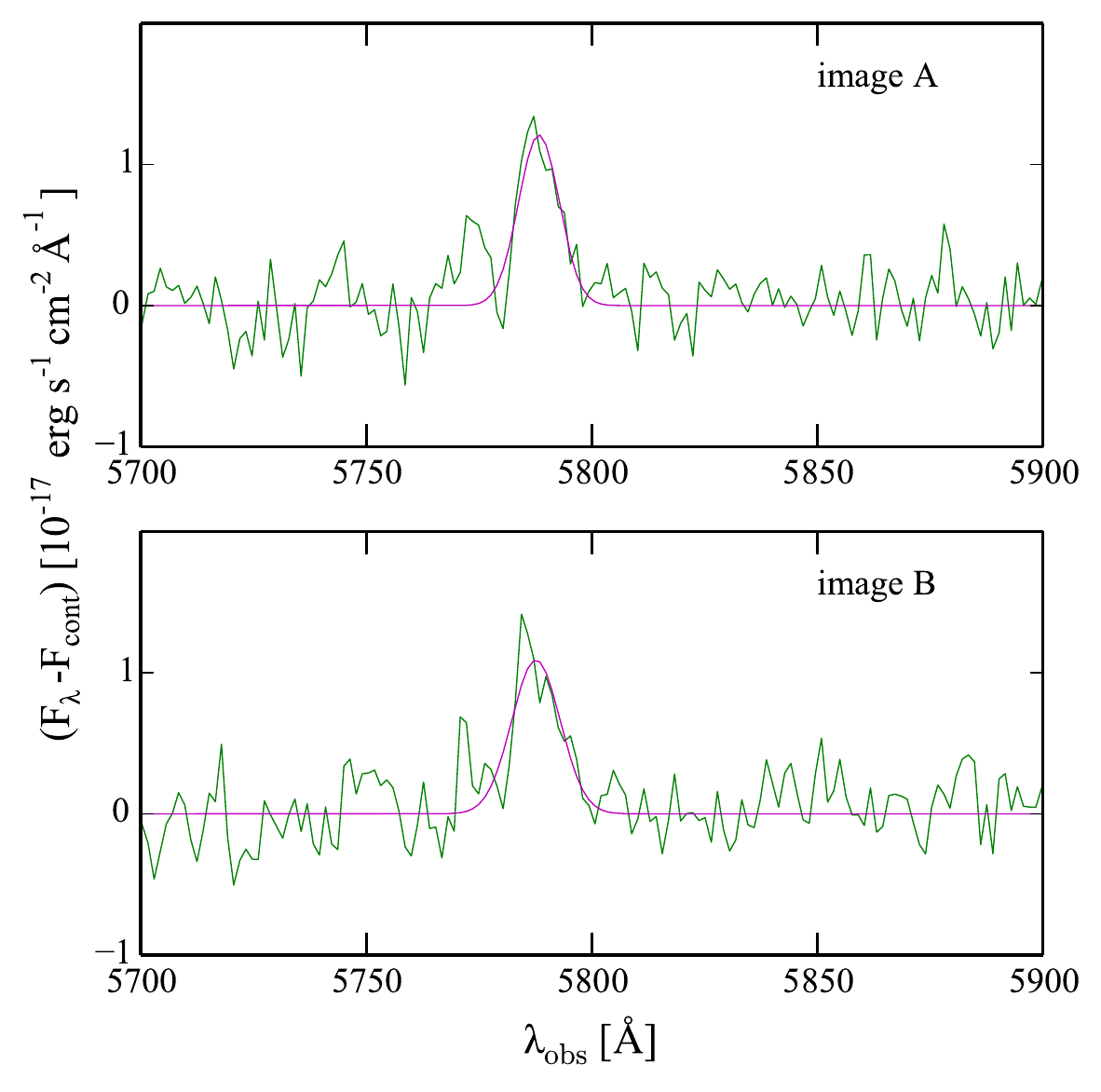}
\end{center}
\caption{\label{fig:gauss} One-dimensional Gaussian fit to the continuum
subtracted spectrum near the Ly-$\alpha$ feature for image A
({\it upper}) and image B ({\it lower}).}

\end{figure}

We fit a Gaussian to the one-dimensional spectrum around the Ly-$\alpha$
emission in images A and B (see \fref{fig:gauss}). We obtain a line
width of $\sigma=4.6\pm0.3$\AA~and $5.4\pm0.4$\AA~at 5788\AA,
respectively. Given the spectral resolution ($R=1918$), the instrumental
broadening corresponds to 156~km~s$^{-1}$. Thus, the Full-Width at
Half-Maximum (FWHM) of Ly-$\alpha$ emission from images A and B, after
accounting for the instrumental broadening, is 540$\pm$40~km~s$^{-1}$
and 640$\pm$50~km~s$^{-1}$, respectively.  There is a second peak
blueward of the Ly-$\alpha$ line, which suggests that the underlying
Ly-$\alpha$ is much broader, with different parts of the emission line
modified by absorption or resonant scattering from surrounding
neutral hydrogen \citep[e.g.][]{MiraldaEscude1998}. This absorption
feature is more obvious for image A than image B. The FWHMs of the
broader Ly-$\alpha$ line (including the second peak to the left) of
images A and B are $\sim 920$ and $\sim 890$~km~s$^{-1}$, respectively.
This is also consistent with the FWHM estimate measured from the stacked
(unbinned) spectra of the four images shown in \fref{fig:spec}.

The flux of the Ly-$\alpha$ line from a flux-calibrated spectrum of
image A is $1.4\times10^{-16}$~ergs~s$^{-1}$~cm$^{-2}$ which gives a
luminosity of $10^{42.5}$~ergs~s$^{-1}$ after taking into account the
magnification factor $\mu$ of 2.5. The equivalent width (EW) of
Ly-$\alpha$ line is $75\pm5$~\AA~in the observed frame and $16\pm1$~\AA
~in the rest frame for image A. The errors represent 68\%
confidence levels on the posterior distribution of the parameters
sampled with MCMC. We present the measurements for both the narrow and
the broad components of Ly-$\alpha$ in \tref{tab:meas}. 

\subsection{Source Magnitude }
\label{sec:srcmag}
{\sc galfit} model fitting to image A in the $i$-band suggested
$i=22.82$.  Using the magnification factor of $\mu=2.5$, we obtain a
delensed source magnitude of $i=23.84$. At $z_{\rm s}=3.76$, this
corresponds to an absolute magnitude $M_{\rm {1500}}=-22.1$ after
applying a K-correction for the continuum $K_{\rm
corr}=-2.5~(1+\alpha)~$~log~$(1+z_{\rm s}) -
2.5~\alpha$~log~$(1500$\AA$/\lambda_{\rm eff})$  following the
prescription of \citet{Glikman2010} where $\alpha=-0.5$
\citep{Richards2006} for quasars.

Alternatively, we can determine the true source magnitude from the
results of our lens mass modeling by fitting an extended source to the
$i$-band image. The best-fit source magnitude thus obtained, is
$i=23.9\pm0.1$ which corresponds to $M_{\rm {UV}}=-22.0\pm0.1$. 

\subsection{Source size}
\label{sec:srcsize}
We followed the same two approaches as before to estimate the true
source size. {\sc galfit} model fitting suggested a size of $r_{\rm
{eff,l}}=0\farcs12$ where subscript $l$ indicates lensed. The lensing
magnification factor ($\mu=2.5$) is used to scale down the area of the
lensed source to get the true source size. We assume the source is
circular (i.e. area=$\pi~r_{\rm {eff,t}}^2$ where subscript $t$ implies
true) and that the lensing magnification is the same in all directions.
Thus, the true source size is estimated to be $r_{\rm
{eff,t}}=\sqrt{r_{\rm {eff,l}}^2/\mu}= 0\farcs069$. Given the physical
scale at $z_{\rm s}=3.76$ of $7.33~$kpc~arcsec$^{-1}$, the 
source size is $\sim 0.56$~kpc in physical units. 

The best-fit source size $r_{\rm {eff,t}}$ from lens mass modeling is
found to be even smaller, $\sim 0\farcs 028\pm0\farcs 005$, suggesting a
physical source size of $0.20\pm0.04$~kpc.  We trust our latter estimate
more because the former estimate has a larger systematic error owing to
the strong assumptions adopted in our calculation.

\section{Discussion}
\label{sec:disc}

In this section, we compare properties of high redshift galaxies
with the lensed source of HSC~J1152+0047 to decide whether the source
has an AGN or not. We also discuss possible sources of
chromatic variations seen in the flux ratio of the brighter pair of
lensed images.

\subsection{Comparison with high redshift galaxies}
Here, we consider comparison with LAEs and LBGs. LAEs are
star-forming galaxies with faint ultra-violet (UV) continuum and a
prominent Ly-$\alpha$ emission line
\citep[e.g.][]{Malhotra2002,Ouchi2008,Shibuya2012}.

LAEs are known to have large EW$_{{\rm Ly-\alpha}}$ and small velocity
widths $\Delta v_{{\rm Ly-\alpha}}$
\citep[e.g.,][]{Ouchi2008,Hashimoto2013}. For example, rest-frame
EW$_{{\rm Ly-\alpha}}$ for a sample of LAEs at $z\sim3.7$ is $>44$~\AA
\citep{Ouchi2008} much larger than that of our lensed source but
their sample is limited by the selection effects of their narrow-band
data.  Usually LAEs are defined to be high-redshift galaxies with
rest-frame EW$_{{\rm Ly-\alpha}}\ge20$\AA~\citep[e.g.,][]{Ouchi2003}. 
Thus, our source is probably at the border of being an LAE based on the
EW criteria. \citet{Ouchi2010} measured velocity widths of Ly-$\alpha$
for a large sample of over 200 Ly-$\alpha$ emitter (LAEs) at high
redshifts. They find average velocity widths of $260$~km~s$^{-1}$
without significant difference between $z=5.7-6.6$. Assuming that this
holds true at $z\sim4$, the FWHM of Ly-$\alpha$ for the source in
HSC~J1152+0047 is much broader to be an LAE type of galaxy.


The characteristic of LBGs is a sharp drop in the flux bluewards of the
Ly-$\alpha$. We do not see any strong evidence of such a sharp cut-off
in either the spectrum (see \fref{fig:spec}) or the imaging although
its colors may not be inconsistent with an LBG \citep[e.g.,][]{Bouwens2014}. 

\citet{Shibuya2015} derived the size-magnitude ($M_{\rm {UV}}$) relation
for SFGs and LBGs at $z\sim4$ (as shown in Figure~9 of their paper). Our
source is an outlier for the $z\sim4$ population and does not show
expected size and magnitude properties for either SFGs or LBGs at that
redshift. In fact, it is extremely compact and at the brighter end of
the rest of the population suggesting that the source probably has an
AGN with low-to-moderate luminosity. 

\subsection{Chromatic variations in the flux ratios}
\label{sec:chrom}
\subsubsection{Intrinsic variability due to AGN}
AGNs are known to show intrinsic variability over months to years
whereas galaxies do not have any mechanism to produce variability on
such time scales, except when they host luminous transients such as
supernovae. From the geometrical symmetry of the lensed images in an
Einstein cross and from our mass models, the relative time delay in the
arrival of light rays between the brighter pair of the lensed images is
a few days. The difference of 1 magnitude in the brighter pair 
(\fref{fig:fxrat}), which are separated by a relative time delay of a
few days, is too rapid to be caused by intrinsic flux variation in the
AGN. 

\subsubsection{Microlensing}
Microlensing is the lensing effect produced by very low mass compact
components, for example, from stars, globular clusters or black holes in
the plane of a lens galaxy. Microlensing could produce chromatic
variations in flux ratios. This effect is sensitive to the scale
of the source size.  The size of region emitting rest-frame optical
(observed frame NIR) in the source plane is much larger than that of the
region emitting rest-frame UV (observed frame optical). Thus, the
$K$s-band fluxes are expected to be less affected by
microlensing. However, HSC~J1152+0047 shows a discrepancy in the
$K$s-band flux ratio instead of the optical. 

Also, the VIKING~$Z$-band (not shown here) suggests a flux ratio,
B/A$\sim 1$ consistent with HSC data. The VIKING~$K$s-band data were
taken 1~yr after VIKING~$Z$-band data and HSC data were taken 4~yrs
after $K$s-band data. Ignoring the fact that we are comparing data in
different bands here, these timescales even though short are still
feasible for microlensing to appear, based on the caustic crossing time
estimated by following \cite{Treyer2004}. 

Microlensing can also be deduced by comparing the continuum and emission
line flux ratios \citep[e.g.][]{sluse2007}.  The ratio of the flux
calibrated optical spectra of images A and B is nearly one, both for the
continuum and the Ly-$\alpha$ emission line disfavoring the microlensing
scenario although the weaker NV and CIV emission lines have different
relative strengths but are not reliable. In summary, microlensing cannot
be considered as the source of the discrepancy between the optical and
NIR flux ratios without additional data. High resolution and deeper
spectroscopy is needed to rule out or confirm microlensing with
certainty. 


\subsubsection{Differential Extinction}
Extinction from the presence of dust in the lens galaxy could affect the
flux ratios as a function of the wavelength. Firstly, the lens galaxy is
an early-type galaxy which usually do not have
significant dust to cause severe extinction of the lensed images
\citep[e.g.][]{Falco1999,Eliasdottir2006}. Secondly, we do not see
any obvious dust lanes in the deep HSC imaging. Given that the system
is fairly symmetrical in the near North-South direction and that images
A and B are almost equidistant on the either side of the lens galaxy,
the amount of dust is not expected to be significantly different to
cause such a high level of differential extinction in the lensed images
A and B and in the $K$s-band alone.  Also, extinction affects strongly
at shorter wavelengths but the flux ratio, B/A does not show any sign of
extinction in the HSC data. Furthermore, the measured flux ratios
in the optical are completely consistent with a simple lens mass model
without the need to include any model for the dust extinction. Thus,
extinction is unlikely to be the source of chromatic variation seen in
HSC~J1152+0047. 

\section{Conclusion}
\label{sec:conc}
We have discovered a quadruply imaged source from the HSC survey at
$z_{\rm s}=3.76$ lensed by an early-type galaxy at $z_{\rm l}=0.466$. 
In this paper, we attempt to understand the whether of the background
source has an AGN at its core by considering various observational evidences.

\begin{itemize}

\item The Ly-$\alpha$ emission line is narrow but broader than
velocity widths typically found in LAEs and LBGs and its luminosity is
consistent with a faint AGN. The rest-frame UV spectrum shows weak metal
emission lines (NV and blue-wing absorbed C~IV) found in the NLR/BLR
around AGNs. 

\item The rest-frame UV absolute magnitude ($M_{UV}=-22$) of the source suggests
that it either belongs to the rare population of intrinsically bright
galaxy population \citep[e.g.,][]{Kobayashi2010,Parsa2016} or the very
abundant population of faint AGNs \citep[e.g.,][]{Glikman2011}. 

\item The source appears point-like in all bands except $i$-band where
only the brighter pair of lensed images suggest extendedness in the
tangential direction. This could suggest that we are detecting emission
from the underlying AGN host galaxy. Alternatively, if there is no AGN,
then the source (with r$_{\rm eff}\sim0.2$~kpc) is one of the most
compact galaxies at $z\sim4$.

\item We do not see any sign of variability in any of the four lensed
images from the HSC data and VIKING $Z$ and $Y$-bands except the VIKING
$K$s-band data which show highly discrepant flux ratio for the brighter
pair. The origin of the chromatic variation is not understood since
none of the known phenomena such as intrinsic variability, microlensing
and differential extinction can fully explain the measured optical and
NIR flux ratios. We can not rule out microlensing due to lack of
sufficient data. 

\end{itemize}
In conclusion, the lensed source is most probably a low-luminosity AGN
(LLAGN) but we can not exclude the possibility of it being a rare, highly
compact bright galaxy.  

If the source indeed has an AGN, then this represents the highest
redshift quadruply lensed AGN. Conventionally, quasars (or bright AGNs)
are favorable for cosmological studies as they are easier to follow-up
on small telescopes for time-delay monitoring. However, LLAGNs also show
variability, in fact, at much shorter time scales and the high redshift,
quadruply lensed source of HSC~J1152+0047 shows great promise for
cosmological studies. Faint AGNs are also useful to study quasar fueling
lifetime, feedback and faint end slope of the luminosity function
\citep[e.g.,][]{Hopkins2006,Glikman2010,Fiore2012,Ikeda2012,Matsuoka2016}.
Studying LLAGNs at high redshifts has been difficult due to their
faintness and strong lensing might be the best means to study inner
structures and mechanisms that drive the nuclear activity in distant LLAGNs. 

Most of the surveys in the past have been shallow and wide enough to
detect numerous lensed quasars (the brighter end of the AGN luminosity
function). The discovery of HSC~J1152+0047 from early HSC data
demonstrates that HSC, with its unique combination of depth and wide
imaging, is beginning to probe the fainter end of the AGN luminosity
function. Looking ahead, we should expect to discover more lensed LLAGNs
from HSC which will open a new avenue of studies. In the future, with
even larger surveys such as the Large Synoptic Survey Telescope survey,
lensed LLAGNs will probably be discovered routinely through variability.

\section*{Acknowledgments}
AM is supported by World Premier International Research Center
Initiative (WPI Initiative), MEXT, Japan. MO is supported by JSPS
KAKENHI Grant Number 26800093 and 15H05892. S.H.S.~is supported by the
Max Planck Society through the Max Planck Research Group.  S.H.S.~and
J.H.H.C.~acknowledge support from the Ministry of Science and Technology
in Taiwan via grant MOST-103-2112-M-001-003-MY3. A.S. acknowledges
support by JSPS KAKENHI Grant Number 26800098. AM would like to thank
T. Anguita, L. Hao, Y. Harikane, S. Huang, D. Sluse and M. Strauss for
useful discussion. The authors would like to thank the referee for
improving the presentation of the paper.
The Hyper Suprime-Cam (HSC) collaboration includes the astronomical
communities of Japan and Taiwan, and Princeton University.  The HSC
instrumentation and software were developed by the National
Astronomical Observatory of Japan (NAOJ), the Kavli Institute for the
Physics and Mathematics of the Universe (Kavli IPMU), the University
of Tokyo, the High Energy Accelerator Research Organization (KEK), the
Academia Sinica Institute for Astronomy and Astrophysics in Taiwan
(ASIAA), and Princeton University.  Funding was contributed by the
FIRST program from Japanese Cabinet Office, the Ministry of Education,
Culture, Sports, Science and Technology (MEXT), the Japan Society for
the Promotion of Science (JSPS),  Japan Science and Technology Agency
(JST),  the Toray Science  Foundation, NAOJ, Kavli IPMU, KEK, ASIAA,
and Princeton University. 
This paper makes use of software developed for the Large Synoptic Survey
Telescope. We thank the LSST Project for making their code available as
free software at \texttt{http://dm.lsstcorp.org}.
The Pan-STARRS1 Surveys (PS1) have been made possible through contributions of the Institute for Astronomy, the University of Hawaii, the Pan-STARRS Project Office, the Max-Planck Society and its participating institutes, the Max Planck Institute for Astronomy, Heidelberg and the Max Planck Institute for Extraterrestrial Physics, Garching, The Johns Hopkins University, Durham University, the University of Edinburgh, Queen's University Belfast, the Harvard-Smithsonian Center for Astrophysics, the Las Cumbres Observatory Global Telescope Network Incorporated, the National Central University of Taiwan, the Space Telescope Science Institute, the National Aeronautics and Space Administration under Grant No. NNX08AR22G issued through the Planetary Science Division of the NASA Science Mission Directorate, the National Science Foundation under Grant No. AST-1238877, the University of Maryland, and Eotvos Lorand University (ELTE).
Based on observations obtained at the Gemini Observatory, which is
operated by the Association of Universities for Research in Astronomy,
Inc., under a cooperative agreement with the NSF on behalf of the Gemini
partnership: the National Science Foundation (United States), the
National Research Council (Canada), CONICYT (Chile), Ministerio de
Ciencia, Tecnolog\'{i}a e Innovaci\'{o}n Productiva (Argentina), and
Minist\'{e}rio da Ci\^{e}ncia, Tecnologia e Inova\c{c}\~{a}o (Brazil).
The authors wish to recognize and acknowledge the very significant
cultural role and reverence that the summit of Mauna Kea has always had
within the indigenous Hawaiian community.  We are most fortunate to have
the opportunity to conduct observations from this mountain.
This research has made use of the NASA/ IPAC Infrared Science Archive,
which is operated by the Jet Propulsion Laboratory, California Institute
of Technology, under contract with the National Aeronautics and Space
Administration.

\bibliographystyle{apj}
\bibliography{references_papers}

\end{document}